\documentclass[twocolumn,aps,prl,showpacs,footinbib]{revtex4}%
\usepackage{graphicx}
\usepackage{amsmath}
\usepackage{amsfonts}
\usepackage{hyperref}
\usepackage{amssymb}%
\def\be{\begin{equation}}
\def\ee{\end{equation}}
\begin{document}
\title{Chaos-induced transparency in an ultrahigh-Q optical microcavity}
\author{Yun-Feng Xiao}
\email{yfxiao@pku.edu.cn}
\author{Xue-Feng Jiang}
\author{Qi-Fan Yang}
\author{Li Wang}
\author{Kebin Shi}
\author{Yan Li}
\author{Qihuang Gong}
\email{qhgong@pku.edu.cn}
\affiliation{State Key Lab for Mesoscopic Physics, Department of Physics, Peking
University, P. R. China}
\date{\today}

\begin{abstract}
We demonstrate experimentally a new form of induced transparency, i.e.,
chaos-induced transparency, in a slightly deformed microcavity which support
both continuous chaotic modes and discrete regular modes with Q factors
exceeding $3\times10^{7}$. When excited by a focused laser beam, the induced
transparency in the transmission spectrum originates from the destructive
interference of two parallel optical pathways: (i) directly refractive
excitation of the chaotic modes, and (ii) excitation of the ultra-high-Q
regular mode via chaos-assisted dynamical tunneling mechanism coupling back to
the chaotic modes. By controlling the focal position of the laser beam, the
induced transparency experiences a highly tunable Fano-like asymmetric
lineshape. The experimental results are modeled by a quantum scattering theory
and show excellent agreement. This chaos-induced transparency is accompanied
by extremely steep normal dispersion, and may open up new possibilities a
dramatic slow light behavior and a significant enhancement of nonlinear interactions.

\end{abstract}

\maketitle

Over the past two decades, it is well known that optical properties of matter
can be dramatically modified by using a secondary light beam. For instance, an
opaque atomic medium is made transparent in the presence of a strong control
beam, known as the unique phenomenon of electromagnetically induced
transparency (EIT) \cite{Boller}. EIT can be explained in terms of a dark
superposition state, or alternatively, by destructive quantum interference of
the transition probability amplitudes. The observation of nonabsorbing
resonance accompanying with the extremely steep normal dispersion through
atomic coherence represents a key feature of EIT, which has led to novel
concepts and important consequences, such as freezing light, enhancing
nonlinear interaction and lasing without inversion \cite{RMP2005}. These have
thrust EIT to the forefront of experimental study in atomic physics during the
last two decades.

It has been recently recognized and demonstrated that similar interference
effects also occur in linear classical systems such as plasma
\cite{Harries1996,Gad}, electric circuits \cite{Litvak,Alzar}, photonic
microresonators
\cite{Smith2004,Maleki2004,Yanik2004,Poon2004,Xu2006,Totsuka2007,Yang2009,Xiao2012,Di2011}%
, various metamaterials
\cite{Zhang2008,Zheludev,Tassin,Liu2009,Kekatpure,Kurter,Hatice,Fan2012,Chen}
and optomechanical systems \cite{Painter,Kippenberg}, which bring the original
quantum phenomena into the realm of classical optics. Remarkably, this
all-optical form of induced transparency does not require naturally occurring
resonances and could therefore be applied to previously inaccessible
wavelength regions, and equally importantly, no strong pumping is necessary.
With a dynamic control in photonic structures, the all-optical EIT even stores
light on a chip at room temperature by breaking the delay-bandwidth limit
\cite{Xu2007}. In this Letter, we demonstrate experimentally a new form of
induced transparency in a slightly deformed optical microcavity on a silicon
chip. The induced transparency originates from the chaos-assisted tunneling
mechanism in the deformed cavity, and thus it is termed as
\textit{chaos-induced transparency}. This tunneling violates the classical law
of ray reflection and represents a formal analogue to the dynamical tunneling
known as a pure quantum mechanical phenomenon \cite{Davis1981}.

\begin{figure}[b]
\begin{center}
\includegraphics[keepaspectratio=true,width=0.45\textwidth]{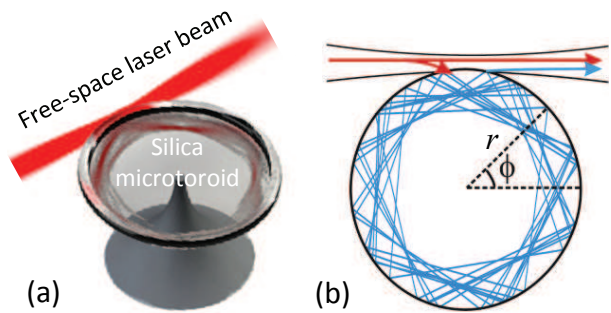}
\end{center}
\caption{(Color online). (a) Schematic illustration of a deformed silica
toroidal microcavity excited by a free-space laser beam. (b) The top view of
the coupling system. The blue curves describe real-space excitation-beam
chaotic trajectories in the cavity. $r,\phi$ gives the polar coordinate system
in the cavity plane.}%
\label{fig1}%
\end{figure}

Figure \ref{fig1} shows a schematic illustration of an on-chip deformed silica
microtoroid excited by a free-space focused laser beam. In our experiment, the
deformed microtoroid is fabricated from a $2$-$\mathrm{\mu m}$-thickness
silicon dioxide layer on a silicon wafer, which possesses a pre-designed
boundary and ultra-smooth cavity surface by combining a \textrm{CO}%
$_{\mathrm{2}}$ laser reflow and a two-step dry etching process
\cite{Jiang,Vahala}. The deformed microtoroid has the boundary defined as
$r\left(  \phi\right)  ={r_{0}(1}+\eta%
{\textstyle\sum\nolimits_{i=2,3}}
{{a_{i}}{{\cos}^{i}}\phi)}$ for $\cos\phi\geq0$, and $r\left(  \phi\right)
={r_{0}(1}+\eta{%
{\textstyle\sum\nolimits_{i=2,3}}
{b_{i}}{{\cos}^{i}}\phi)}$ for $\cos\phi<0$ in the polar coordinates ($r$,
$\phi$), where $\eta$ denotes the deformation parameter related to the aspect
ratio of the shape, and ${r_{0}}$ represents the size parameter. In our
experiment, $\eta\sim1$, ${r_{0}}\sim45$ $\mathrm{\mu m}$, $a_{2}\sim-0.1329$,
$a_{3}\sim0.0948$, $b_{2}\sim-0.0642$, and $b_{3}\sim-0.0224$. Other than the
intrinsic isotropy of emission in rotationally symmetric cavities, the
deformed microtoroid supports both ultrahigh Q factors and highly directional
emission toward $180%
\operatorname{{{}^\circ}}%
$ far-field direction (emitted at polar angles $\phi\sim\pi/2$ and $3\pi/2$)
\cite{Jiang,Zou}. The underlying physics of highly directional emission lies
in the dynamical tunneling \cite{Nature1997,Narimanov,Song}.

An important production of the aforementioned directional emission is that we
can directly inject a laser beam into chaotic orbits of the deformed cavity
with a time reversed way, and then the excitation light might be finally
transferred to a target high-Q mode via the chaos-assisted dynamical tunneling
\cite{An,Park}. As the tunneling is a wave-mechanical process, this transfer
process would occur efficiently when the excitation light is on resonance with
the target mode. In our experiment, a tunable laser beam is focused near the
edge of the deformed microtoroid. The focused beam waist is smaller than $3$
$\mathrm{\mu}$\textrm{m} in diameter. The microcavity is mounted on a
rotational stage with $1%
\operatorname{{{}^\circ}}%
$ angular resolution and a translational stage with $20$ \textrm{nm}
resolution, which allow a precise control of the coupling between the
free-space beam and the microcavity by adjusting its incident far-field angle
$\theta$ and the radial displacement $\Delta r$ relative to the local cavity boundary.

\begin{figure}[tb]
\begin{center}
\includegraphics[keepaspectratio=true,width=0.45\textwidth]{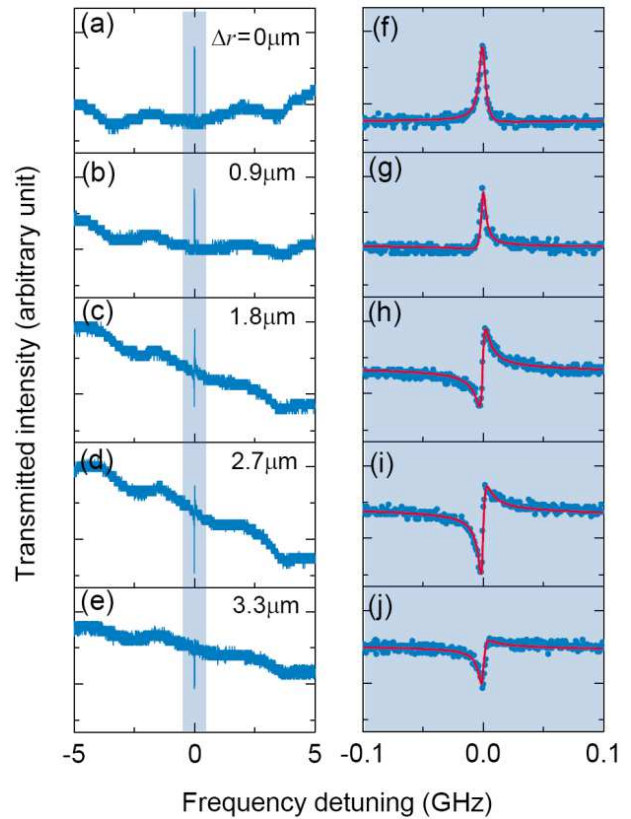}
\end{center}
\caption{(Color online) (a)-(e), Measured transmission spectra depending on
the excitation position of the focused beam for an input power of $10$
$\mathrm{\mu}$\textrm{W}. All figures have the same scales in both horizontal
and vertical axes. From top to bottom, the focused beam moves away from the
cavity boundary where the case of the highest induced transparency peak is set
as the initial position ($\Delta r=0$). (f)-(j) Zoom-in of the transmission
spectra around the transparency window. Red solid curves correspond to model
fits to the experimental data (blue dots).}%
\label{fig2}%
\end{figure}

In the first experiment, the incident beam injects from $180%
\operatorname{{{}^\circ}}%
$ far-field direction and focuses on the cavity edge at $\phi\sim\pi/2$, and
the collection of transmitted light is in $0%
\operatorname{{{}^\circ}}%
$ far-field direction. Figure \ref{fig2}(a) with the zoom-in (f) shows a
typical power transmission spectrum for a quasi-TM mode of the coupling
system. It can be seen that the high-Q mode is eventually excited, and
importantly, the sharp EIT-like resonance occurs. To test the dependence of
the transparency resonance on the separation between the microcavity and the
laser focal spot, transmission spectra are measured by radially moving the
focused beam away from the cavity boundary, as shown in Figs. \ref{fig2}%
(b)-\ref{fig2}(e) and \ref{fig2}(g)-\ref{fig2}(j). Obviously, the EIT-like
resonance experiences a transition to asymmetric Fano resonance \cite{Fano}.
The observation of Fano-like EIT resonance at the ultrahigh-Q cavity mode is
attributed to the chaos-assisted dynamical tunneling in the deformed cavity,
because such a high-Q mode can not be directly excited by a free-space beam
due to the angular momentum mismatching between them.

\begin{figure}[tb]
\begin{center}
\includegraphics[keepaspectratio=true,width=0.45\textwidth]{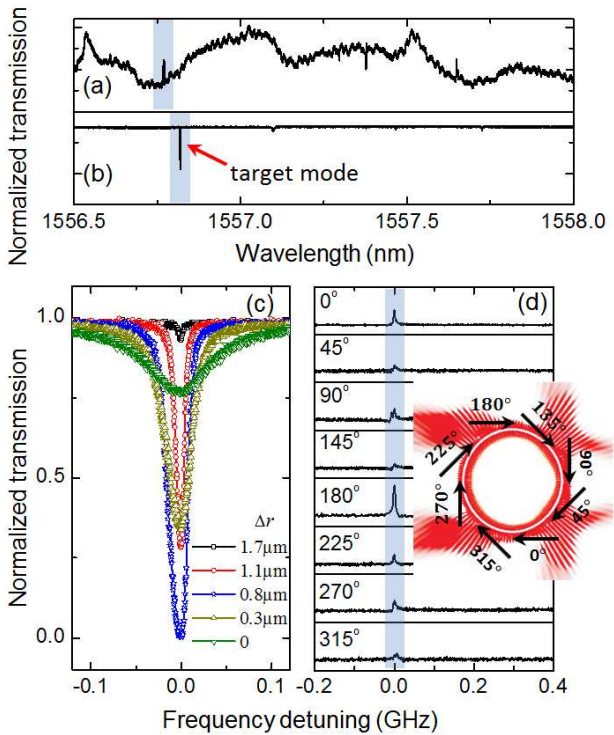}
\end{center}
\caption{(Color) Comparison of the transmission spectra obtained with the
focused beam excitation (a) and the fiber taper coupling (b). All modes
observed in (a) can find their correspondences in (b) where a minor red shift
occurs when coupled to the taper. The red arrow indicates the
ultra-high-Q\ target mode ($Q\sim3\times10^{7}$) studied throughout the paper.
(c) The transmission spectra around the same mode when coupled to the taper
for different taper-cavity gaps. The blue curve shows the critical coupling.
Black and red curves correspond to the under-coupled cases while yellow and
green curves show the over-coupled cases. (d) The transmission spectra around
the target mode for eight typical coupling positions of the focused beam shown
in the inset. Here each spectrum is obtained to have the highest transparency
peak by optimizing the coupling position.}%
\label{fig3}%
\end{figure}

To further investigate the role of the dynamical tunneling, in the second
experiment, we use a fiber taper (waist diameter $\sim1.5$ $\mathrm{\mu}%
$\textrm{m}) to couple the deformed cavity. Figures \ref{fig3}(a)-\ref{fig3}%
(b) compare the transmission spectra obtained with the two types of coupling
methods. It can be seen that high-Q modes in both the spectra show a good
correspondence where a minor red shift of modes occurs when coupled to the
taper. The mode with the intrinsic quality factor exceeding $3\times10^{7}$ at
the wavelength of $1556.8$ \textrm{nm} is the target mode throughout this
paper. Remarkably, the EIT-like characteristic in the free-space transmission
spectrum differs greatly from the symmetric Lorenzian dips (the dips are due
to the cavity loss in the taper coupling). Moreover, no obvious change of the
resonance lineshape is observed when the taper-cavity system is tuned
continuously from under- to deep over-coupled regions, as depicted in Fig.
\ref{fig3}(c). This is because that the taper directly excites high-Q modes
(such as whispering gallery modes) instead of chaotic modes thanks to the
angular momentum matching \cite{Vahala}, and the dynamical tunneling between
them essentially contributes to an additional energy decay for the high-Q
modes in the present case. Even for the free-space coupling to the deformed
cavity, the transparency resonances strongly depend on the excitation
positions. As shown in Fig. \ref{fig3}(d), the highest transparency peak
occurs when the incident beam focuses on the cavity edge at $\phi\sim\pi/2$
and from $180%
\operatorname{{{}^\circ}}%
$ far-field direction. This is the exact position where high-Q
counter-clockwise modes in the deformed cavity shows the strongest universal
directional emission assisted by the chaos. In our experiment, we also
measured the transmission spectra of an undeformed toroidal microcavity, and
no Fano-like EIT resonance is observed with both the coupling methods because
no chaotic mode is supported in the undeformed cavity.

\begin{figure}[tb]
\begin{center}
\includegraphics[keepaspectratio=true,width=0.45\textwidth]{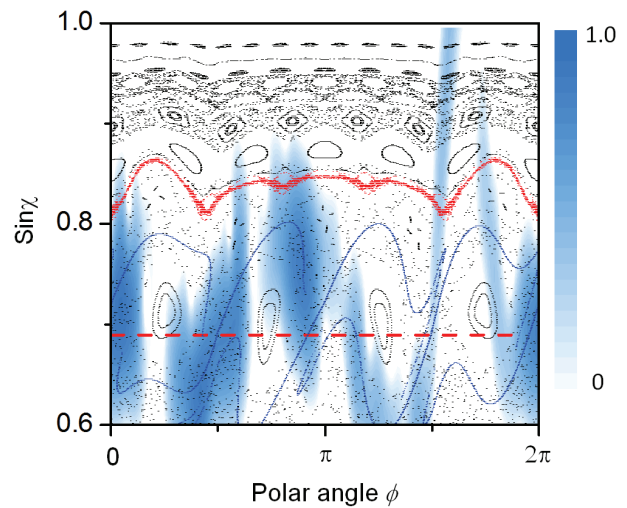}
\end{center}
\caption{(Color) Phase-space structure (Poincar\'{e} surface of section) for
the deformed microcavity plotted in the Birkhoff's coordinates, in which a ray
is reflected off the cavity boundary at polar angle $\phi$\ with an incident
angle $\chi$. The cavity parameters are defined in the text. The red dotted
curve above $\sin\chi=0.8$ stands for a KAM torus, which divides the chaotic
sea into classically accessible (bottom) and classically forbidden (upper)
regions. The red dashed curve at $\sin\chi=0.69$ indicates the critical
refraction line. Husimi distribution (logarithmic scale) of the excitation
field is shown in sky-blue. Blue curves below the red KAM torus represent the
excited-chaos trajectories.}%
\label{fig4}%
\end{figure}

The physical mechanism of chaos-induced transparency is studied briefly in the
following. The features of internal ray dynamics within the deformed
microcavity can be well described in the phase space (Poincar\'{e} surface of
section) \cite{Nature1997,Narimanov}, as shown in black in Fig. \ref{fig4}. In
this phase-space structure, ray dynamics is mostly chaotic, in addition to
some regular orbits such as islands and Kalmorogov-Arnol'd-Moser (KAM) tori
\cite{KAM}. Classically, these different structures are disjoint. For
instance, chaotic rays below the red KAM torus cannot couple into high-Q modes
which are typically located on the upper of the phase space. This is verified
by the chaotic trajectories shown as blue curves in Fig. \ref{fig4}, which
cannot cross the KAM torus. However, in reality, the dynamical tunneling
between the regular modes and neighboring chaotic orbits takes place. To
demonstrate this, the sky-blue shadow in Fig. \ref{fig4} plots the Husimi
projection of the excited chaotic modes, representing the wave analog of the
phase space. As expected, the most excitation field is distributing in the
chaos region below the KAM torus, because the free-space beam refracts into
the cavity and directly excites chaotic modes. Remarkably, some tails
obviously intrude into the upper area by crossing the KAM torus. In
particular, a strong tail at $\phi\sim$ $3\pi/2$ even reaches the top of the
phase space where ultrahigh-Q modes are located. In other words, these
ultrahigh-Q modes can be eventually excited via the chaos-assisted tunneling.
As for the Fano-like EIT resonances, there exists two parallel excitation
pathways: (1) direct excitation of the continuous chaotic modes from the
incident beam, and (2) excitation of the high-Q mode via the chaos-assisted
tunneling coupling back to the chaotic modes. These two pathways interfere
with each other in the far field because a phase shift occurs when light
crosses KAM tori which represents a potential barrier in wave-optic field.

We now theoretically model the experimental results in more details. In the
free-space coupling process shown in Fig. \ref{fig1}, the incident light
$|\mathrm{in}\rangle$ with frequency $\omega$\ is scattered by the
microcavity, and directly excites continuous chaotic modes, denoting the
excitation state $|\mathrm{C}_{\omega}\rangle$. We consider that a discrete
regular mode $|\mathrm{M}\rangle$ with resonant frequency $\omega_{0}$
interacts with the chaotic states, and assume they are orthogonal \cite{An}.
The interacting system is governed by a Hamiltonian $H$, satisfying
$\langle\mathrm{M}|H|\mathrm{M}\rangle=\omega_{0}-i\gamma/2$, $\langle
\mathrm{C}_{\omega}|H|\mathrm{M}\rangle=V_{\omega}$, and $\langle
\mathrm{C}_{\omega^{\prime}}|H|\mathrm{C}_{\omega}\rangle=\omega\delta
(\omega^{\prime}-\omega)$. Here $V_{\omega}$ describes the interaction between
$|\mathrm{M}\rangle$ and $|\mathrm{C}_{\omega}\rangle$, known as the dynamical
tunneling; $\gamma$ represents the modified decay rate of $|\mathrm{M}\rangle
$, consisting of (i) the intrinsic loss such as radiation, material absorption
and scattering, and (ii) the chaos-assisted tunneling into the chaos other
than the excitation state.

With a standard procedure developed by Fano in Ref. \cite{Fano}, we obtain the
transmission spectrum of the free-space beam
\begin{equation}
T(\omega)=\frac{|q_{\omega}+\epsilon-iK|^{2}}{(1+K)^{2}+\epsilon^{2}}%
|\langle\mathrm{C}_{\omega}|S|\mathrm{in}\rangle|^{2}. \label{tran}%
\end{equation}
Here $K$ is defined as the ratio $\gamma/\kappa\equiv\left(  \gamma_{t}%
-\kappa\right)  /\kappa$ with $\kappa=2\pi|V_{\omega}|^{2}$ being the coupling
strength and $\gamma_{t}\equiv\kappa+\gamma$ representing the whole decay rate
of the regular mode, where $V_{\omega}$ remains constant under the first
Markov approximation \cite{QN}; $\epsilon\equiv\left(  \omega-\omega
_{0}\right)  /(\kappa/2)$ describes the normalized detuning between the
incident light and the regular mode; $S$ is a suitable transition operator
between $|\mathrm{in}\rangle$ and $|\mathrm{C}_{\omega}\rangle$, and
$|\langle\mathrm{C}_{\omega}|S|\mathrm{in}\rangle|^{2}$ describes the
probability of transmitted signal \cite{Fano}; $q_{\omega}=\langle
\varphi_{\omega}|S|\mathrm{in}\rangle/\left(  \pi V_{\omega}^{\ast}%
\langle\mathrm{C}_{\omega}|S|\mathrm{in}\rangle\right)  $ stands for the shape
parameter of the transmission spectrum, where $\left\vert \varphi_{\omega
}\right\rangle =|\mathrm{M}\rangle+{\mathcal{P}}\int\mathrm{d}\omega^{\prime
}V_{\omega^{\prime}}/\left(  \omega-\omega^{\prime}\right)  \left\vert
\mathrm{C}_{\omega^{\prime}}\right\rangle $ with $\mathcal{P}$ denoting
Cauchy's principle value. To give a clear understanding, we consider two
special cases as follows.

(i) If the dynamical tunneling between the regular and chaotic modes is
absent, \textit{i.e.}, $\kappa=0$, the transmission spectrum $T(\omega)$ in
the absence of high-Q regular mode reduces to $T_{0}(\omega)=|\langle
\mathrm{C}_{\omega}|S|\mathrm{in}\rangle|^{2}$, known as unperturbed
scattering due to $\epsilon,K\rightarrow\infty$ in this case.

(ii) If the regular mode is over-coupled, \textit{i.e.}, $\gamma\ll\kappa$ and
$K\rightarrow0$, the transmission spectrum yields a standard Fano resonance
$T_{f}(\omega)=|q_{\omega}+\epsilon|^{2}/\left(  1+\epsilon^{2}\right)
T_{0}(\omega)$.

In experiments, the unperturbed transmission spectrum $T_{0}(\omega)$ shows
baseline oscillations in a large scanning range, similar to Refs.
\cite{Poon2004,Yang2010}, which can be simply modeled by the interference of
two amplitudes: (i) the directly transmitted amplitude $t$, and (ii) the
dissipated amplitude $r$ that refracts into and back from the cavity with an
additional phase shift $\theta$. Thus, the unperturbed transmission reads
$T_{0}(\omega)=\left\vert t+re^{i\theta}\right\vert ^{2}$, and the shape
parameter is simplified, given by $q_{\omega}=-ire^{i\theta}/\left(
t+re^{i\theta}\right)  $, because the contribution of $\langle\mathrm{M}%
|S|\mathrm{in}\rangle$ is negligible for a high-Q regular mode. As both
$T_{0}(\omega)$ and $q_{\omega}$ are determined, we can finally obtain the
transmission spectrum $T(\omega)$ in Eq. \ref{tran}.

\begin{figure}[tb]
\begin{center}
\includegraphics[keepaspectratio=true,width=0.4\textwidth]{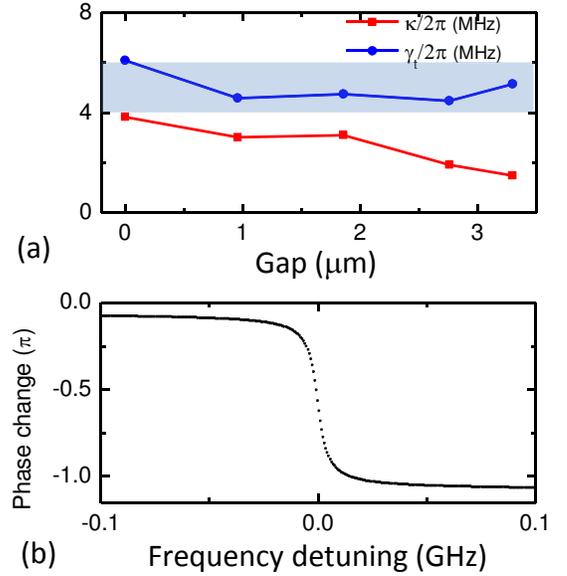}
\end{center}
\caption{(Color online) (a) Modeled $\kappa$ and $\gamma_{t}$ in Figs.
2(f)-2(j) depending on the beam-cavity gap. The shadow shows the measured
linewidth range with a fiber taper coupling. (b) The calculated phase change
of the transmitted field in Fig. 2(f), showing a strong normal dispersion.}%
\label{fig5}%
\end{figure}

Under this model, the red solid curves in Figs. \ref{fig2}(f)-\ref{fig2}(j)
show the theoretical fittings, in good accordance with the experimental
spectra (blue dots). From top to bottom, the transmission spectra experience a
transition from the symmetric induced transparency to asymmetric line shape,
predominantly due to the additional phase shift $\theta$ varying from $\pi$ to
$2\pi$\textit{ }during this process. The fitting parameters $\kappa/2\pi$ and
$\gamma_{t}/2\pi$ are plotted in Fig. \ref{fig5}(a). It can be seen that the
coupling between the regular and the excitation states, described by the
strength $\kappa$, becomes weaker when the focused beam moves away from the
cavity boundary. This is possibly because that the direct refraction is less
and the excited $|\mathrm{C}_{\omega}\rangle$\ tends to distributes lower in
the phase space. Furthermore, the modeled total decay rates $\gamma_{t}/2\pi$
of the regular mode range from $4$ to $6$ \textrm{MHz}, representing an
invariant in our theory, which exactly fall within the error of the resonant
linewidth measured by using the taper coupling method.

To characterize the chaos-induced transparency, finally, Fig. \ref{fig5}(b)
shows the calculated phase change of the transmitted field with the modeled
data in Fig. \ref{fig2}(f). This conventional curve exhibits an extremely
steep normal dispersion, and indicates a strong suppression of the group
velocity. At the resonance point $\omega\sim\omega_{0}$, the group velocity is
significantly reduced to smaller than $10^{-5}c$ where $c$ represents the
velocity of light in vacuum, resulting from the ultra-narrow linewidth of the
ultra-high-Q mode in the deformed microcavity.

In summary, we have experimentally demonstrated the chaos-induced transparency
in a deformed microcavity with Q factor exceeding $3\times10^{7}$. A
theoretical model is present and agrees well with the experimental results.
When excited by a focused laser beam, the induced transparency in the
transmission spectrum is attributed to the destructive interference of two
parallel optical pathways either resonantly exciting the high-Q mode through
the chaos-assisted tunneling or not. By controlling the excitation position of
the laser beam, the induced transparency experiences a highly tunable
Fano-like asymmetric lineshape. This chaos-induced transparency is accompanied
by the extremely steep normal dispersion resulting from the ultrahigh-Q mode,
which provides a dramatic slow light behavior and a significant enhancement of
nonlinear interactions in the optical microcavity.

\begin{acknowledgments}
Y.F.X. thanks Professor Kyungwon An at Seoul National University for stimulating
discussions and suggestions .This work was supported by the National Basic Research Program of China (No.
2007CB307001), National Natural Science Foundation of China (Nos. 11004003,
11121091, and 11023003), and the Research Fund for the Doctoral Program of
Higher Education of China (No. 20090001120004).
\end{acknowledgments}


\begin{thebibliography}{99}                                                                                               %


\bibitem {Boller}K. J. Boller, A. Imamoglu, and S. E. Harris, \textrm{Phys.
Rev. Lett.} \textbf{66}, 2593 (1991).

\bibitem {RMP2005}M. Fleischhauer, A. Imamoglu, and J. P. Marangos,
\textrm{Rev. Mod. Phys.} \textbf{77}, 633 (2005).

\bibitem {Harries1996}S. E. Harris, \textrm{Phys. Rev. Lett.} \textbf{77},
5357 (1996).

\bibitem {Gad}R. Gad, J. G. Leopold, A. Fisher, D. R. Fredkin, and A. Ron,
\textrm{Phys. Rev. Lett.} \textbf{108}, 155003 (2012).

\bibitem {Litvak}A. G. Litvak and M. D. Tokman, \textrm{Phys. Rev. Lett.}
\textbf{88}, 095003 (2002).

\bibitem {Alzar}C. L. G. Alzar, M. A. G. Martinez, and P. Nussenzveig,
\textrm{Am. J. Phys.} \textbf{70}, 37 (2002).

\bibitem {Smith2004}D. D. Smith, H. Chang, K. A. Fuller, A. T. Rosenberger,
and R. W. Boyd, \textrm{Phys. Rev. A} \textbf{69}, 063804 (2004).

\bibitem {Maleki2004}L. Maleki, A. B. Matsko, A. A. Savchenkov, and V. S.
Ilchenko, \textrm{Opt. Lett.} \textbf{29}, 626 (2004).

\bibitem {Yanik2004}M. F. Yanik, W. Suh, Z. Wang, and S. Fan, \textrm{Phys.
Rev. Lett.} \textbf{93}, 233903 (2004).

\bibitem {Poon2004}H.-T. Lee and A. W. Poon, \textrm{Opt. Lett.} \textbf{29},
5 (2004).

\bibitem {Xu2006}Q. Xu, S. Sandhu, M. L. Povinelli, J. Shakya, S. Fan, and M.
Lipson, \textrm{Phys. Rev. Lett.} \textbf{96}, 123901 (2006).

\bibitem {Totsuka2007}K. Totsuka, N. Kobayashi, and M. Tomita, \textrm{Phys.
Rev. Lett.} \textbf{98}, 213904 (2007).

\bibitem {Yang2009}X. Yang, M. Yu, D.-L. Kwong, and C. W. Wong, \textrm{Phys.
Rev. Lett.} \textbf{102}, 173902 (2009).

\bibitem {Xiao2012}B.-B. Li, Y.-F. Xiao, C.-L. Zou, X.-F. Jiang, Y.-C. Liu,
F.-W. Sun, Y. Li, and Q. Gong, \textrm{Appl. Phys. Lett.} \textbf{100}, 021108 (2012).

\bibitem {Di2011}K. Di, C. Xie, and J. Zhang, \textrm{Phys. Rev. Lett.}
\textbf{106}, 153602 (2011).

\bibitem {Zhang2008}S. Zhang, D. A. Genov, Y. Wang, M. Liu, and X. Zhang,
\textrm{Phys. Rev. Lett.} \textbf{101}, 047401 (2008).

\bibitem {Zheludev}N. Papasimakis, V. A. Fedotov, N. I. Zheludev, and S. L.
Prosvirnin, \textrm{Phys. Rev. Lett.} \textbf{101}, 253903 (2008).

\bibitem {Tassin}P. Tassin, L. Zhang, Th. Koschny, E. N. Economou, and C. M.
Soukoulis, \textrm{Phys. Rev. Lett.} \textbf{102}, 053901 (2009).

\bibitem {Liu2009}N. Liu, L. Langguth, T. Weiss, J. K\"{a}stel, M.
Fleischhauer, T. Pfau, and H. Giessen, \textrm{Nature Materials} \textbf{8},
758 (2009).

\bibitem {Kekatpure}R. D. Kekatpure, E. S. Barnard, W. Cai, and M. L.
Brongersma, \textrm{Phys. Rev. Lett.} \textbf{104}, 243902 (2010).

\bibitem {Kurter}C. Kurter, P. Tassin, L. Zhang, T. Koschny, A. P. Zhuravel,
A. V. Ustinov, S. M. Anlage, and C. M. Soukoulis, \textrm{Phys. Rev. Lett.}
\textbf{107}, 043901 (2011).

\bibitem {Hatice}A. Artar, A. A. Yanik, and H. Altug, \textrm{Nano Lett.}
\textbf{11}, 1685 (2011).

\bibitem {Fan2012}L. Verslegers, Z. Yu, Z. Ruan, P. B. Catrysse, and S. Fan,
\textrm{Phys. Rev. Lett.} \textbf{108}, 083902 (2012).

\bibitem {Chen}J. Chen, Z. Li, S. Yue, J. Xiao, and Q. Gong, \textrm{Nano
Lett.} \textbf{12}, 2494 (2012).

\bibitem {Painter}A. H. Safavi-Naeini, T. P. Mayer Alegre, J. Chan, M.
Eichenfield, M. Winger, Q. Lin, J. T. Hill, D. E. Chang, and O. Painter,
\textrm{Nature} \textbf{472}, 69 (2011).

\bibitem {Kippenberg}S. Weis, R. Rivi\`{e}re, S. Del\'{e}lise, E. Gavartin, O.
Arcizet, A. Schliesser, and T. J. Kippenberg, \textrm{Science} \textbf{300},
1520 (2010).

\bibitem {Xu2007}Q. Xu, P. Dong, and M. Lipson, \textrm{Nature Phys.}
\textbf{3}, 406 (2007).

\bibitem {Davis1981}M. J. Davis and E. J. Heller, \textrm{J. Chem. Phys.}
\textbf{75}, 246 (1981).

\bibitem {Vahala}D. K. Armani, T. J. Kippenberg, S. M. Spillane, and K. J.
Vahala, \textrm{Nature} \textbf{421}, 925 (2003).

\bibitem {Jiang}X.-F. Jiang, Y.-F. Xiao, C.-L. Zou, L. He, C.-H. Dong, B.-B.
Li, Y. Li, F.-W. Sun, L. Yang, and Q. Gong, \textrm{Adv. Mat.}, in press.

\bibitem {Zou}C.-L. Zou, F.-W. Sun, C.-H. Dong, X.-W. Wu, J.-M. Cui, Y. Yang,
G.-C. Guo, Z.-F. Han, http://arxiv.org/abs/0908.3531.

\bibitem {Nature1997}J. U. N\"{o}ckel and A. D. Stone, \textrm{Nature}
\textbf{385}, 45 (1997).

\bibitem {Narimanov}V. A. Podolskiy and E. E. Narimanov, \textrm{Opt. Lett.}
\textbf{30}, 474 (2005).

\bibitem {Song}Q. Song, L. Ge, B. Redding, and H. Cao, \textrm{Phys. Rev.
Lett.} \textbf{108}, 243902 (2012).

\bibitem {An}J. Yang, S. B. Lee, S. Moon, S. Y. Lee, S. W. Kim, T. T. A. Dao,
J. H. Lee, and K. An, \textrm{Phys. Rev. Lett.} \textbf{104}, 243601 (2010).

\bibitem {Park}Y. S. Park and H. Wang, \textrm{Nature Physics} \textbf{5}, 489 (2009).

\bibitem {Fano}U. Fano, \textrm{Phys. Rev.} \textbf{124}, 1866 (1961).

\bibitem {KAM}V. F. Lazutkin, KAM theory and semiclassical approximations to
eigenfunctions (Springer, New York, 1993).

\bibitem {QN}C.W. Gardiner and P. Zoller, Quantum Noise, 3rd ed. (Springer,
Berlin, 2004).

\bibitem {Yang2010}J. Yang, S. B. Lee, S. Moon, S. Y. Lee, S. W. Kim, and K.
An, \textrm{Opt. Express} \textbf{18}, 26141 (2010).
\end{thebibliography}
\end{document}